\documentclass[journal]{IEEEtran}

\usepackage{amsmath,amssymb,amsfonts}
\usepackage{graphicx}
\usepackage{booktabs}
\usepackage{xcolor}
\usepackage{tikz}
\usetikzlibrary{arrows.meta,positioning,fit,backgrounds,calc,shapes.geometric}
\usepackage{url}
\usepackage{algorithm}
\usepackage{algpseudocode}
\usepackage{cite}
\usepackage[hidelinks]{hyperref}

\usepackage{bbm}

\graphicspath{{figs/}}

\newcommand{\CARLTON}{\textsc{Carlton}}
\newcommand{\OURS}{\textsc{FASCIA}}
\newcommand{\FedAvg}{\textsc{FedAvg}}
\newcommand{\FedProx}{\textsc{FedProx}}

\renewcommand{\baselinestretch}{0.99}

\begin{document}

\title{Federated Agentic AI for Interference-Aware Adaptive Spectrum Coordination}

\author{Tomer~Baziza~Paz,
        Yaniv~Cohen,
        and~Kobi~Cohen~\IEEEmembership{(Senior Member,~IEEE)}
\thanks{The authors are with the School of Electrical and Computer Engineering, Ben-Gurion University of the Negev, Beer-Sheva 8410501, Israel.}
\thanks{Corresponding author: K. Cohen, email: 
yakovsec@bgu.ac.il}
\thanks{This work has been submitted to the IEEE for possible publication.
Copyright may be transferred without notice, after which this version may
no longer be accessible.}
}

\maketitle

%==============================================================================
\begin{abstract}
Emerging intelligent wireless networks increasingly rely on autonomous AI agents to adapt spectrum usage to dynamic and heterogeneous operating conditions. However, collective learning is challenging when agents experience different local interference environments and cannot exchange raw observations due to communication-overhead and privacy constraints. We propose Federated Agentic Spectrum Coordination with Interference Awareness (FASCIA), a federated multi-agent deep reinforcement learning framework for adaptive spectrum coordination. In FASCIA, each network manager acts as an
autonomous AI agent that perceives its local interference environment,
accumulates experience in a persistent local memory, and learns an adaptive spectrum coordination policy through local
interactions. Coordination among agents is achieved through periodic federated aggregation of locally trained models, without exchanging raw sensing data or experience. Since spectrum decisions made by one agent alter the environments observed by others, local learning is inherently heterogeneous and coupled. FASCIA employs proximal regularization to mitigate the resulting model drift and explicitly accounts for inter-carrier interference arising from partially overlapping channels. FASCIA is the first federated agentic AI framework to address this challenging interference-coupled spectrum coordination setting. Extensive simulations demonstrate that FASCIA achieves performance
close to the centralized benchmark while avoiding raw-experience
exchange and maintaining effectiveness across heterogeneous network
configurations.
\end{abstract}


%==============================================================================
\section{Introduction}
\label{sec:intro}
%==============================================================================

The growing demand for wireless connectivity, coupled with spectrum scarcity and the increasing density of independently managed wireless networks, is driving a transition toward more autonomous and adaptive spectrum coordination. This trend is particularly pronounced in emerging 6G edge environments, where multiple networks may coexist within a shared spectrum band while experiencing rapidly changing traffic, topology, and interference conditions. Effective operation in such environments requires more than static resource allocation or one-shot optimization. Network managers must continuously observe their local wireless environment, learn from previous interactions, adapt their spectrum decisions to changing conditions, and coordinate their behavior with other networks. Adaptive spectrum coordination (ASC) captures this closed-loop decision-making process, in which independently managed networks dynamically select and revise their spectrum usage according to evolving interference conditions. Dynamic channel allocation constitutes one important realization of ASC. Such adaptive spectrum management is relevant to cognitive radio networks, 6G in-X subnetworks, vehicular ad-hoc networks, wireless sensor and IoT networks, and wireless mesh networks, where independently managed networks must efficiently coexist over shared spectrum resources~\cite{liu2012cooperative,wang2021decentralized,adeogun2022multi,du2022multi,cohen2024carlton}.

\subsection{Challenges of Agentic Interference-Coupled Adaptive Spectrum Coordination}

A fundamental challenge in ASC is that spectrum interactions are considerably more complex than those captured by conventional orthogonal-channel models. Many existing formulations assume perfectly orthogonal channels and therefore model interference primarily between networks selecting the same channel. In practical wireless systems, however, spectral leakage, imperfect filtering, and other hardware limitations give rise to inter-carrier interference (ICI), such that transmissions over partially overlapping channels can significantly affect one another~\cite{mishra2006overlapped}. Consequently, the interference experienced by a network depends not only on co-channel transmissions but also on activity over neighboring channels, with the impact generally decreasing with spectral separation. Spectrum decisions therefore become interference-coupled: the effect of a local decision depends jointly on spectral separation, network topology, and the simultaneous decisions of other networks. This weighted, non-binary interference structure makes adaptive coordination substantially more challenging than conventional allocation under ideal channel orthogonality.

The problem becomes even more challenging when spectrum coordination must be learned autonomously and executed in a distributed manner. Each network manager (i.e., agent) has access primarily to its own sensing information and therefore possesses only a partial view of the global interference state, yielding a partially observable decision-making problem~\cite{naparstek2018deep,adeogun2022multi,kaur2023deep}. Moreover, the environment experienced by each network is inherently non-stationary: whenever one network modifies its spectrum usage, it changes the interference observed by other networks and consequently affects their observations, rewards, and subsequent decisions. Thus, the agents do not learn in isolated environments; rather, their local learning processes are intrinsically coupled through the shared wireless medium~\cite{du2022multi,adeogun2023distributed}. At the same time, computing the optimal joint spectrum assignment generalizes graph coloring to weighted, non-binary interference and is computationally intractable~\cite{blum2003metaheuristics}, making repeated centralized optimization unsuitable for rapidly evolving network conditions. Finally, adaptation itself incurs a cost, since frequent channel changes interrupt ongoing communication and increase spectrum mobility. Hence, an effective ASC policy must react rapidly to changing interference conditions while minimizing unnecessary spectrum mobility.

Beyond these decision-making challenges, enabling collective learning among autonomous network managers introduces additional communication and privacy concerns. Centralized learning architectures may require locally collected sensing observations and interaction experiences to be transferred to a central entity, resulting in considerable communication overhead and exposing potentially sensitive operational information~\cite{niknam2020federated,lim2020federated,chen2021joint}. These limitations become increasingly important as the number of autonomous networks grows and their local operating conditions become more heterogeneous~\cite{kairouz2021advances,li2020fedprox}. An effective learning architecture should therefore allow individual network managers to learn continuously from their local interactions while benefiting from knowledge acquired across the network, without requiring the exchange of raw sensing observations or interaction experiences.

\subsection{Federated Agentic AI under Interference Coupling}

Federated Agentic AI provides a natural framework for realizing this form of
distributed intelligence. Rather than centralizing the experience collected
throughout the network, autonomous AI agents can maintain and learn from their
local observations and interaction histories while periodically sharing learned
model information to support collective adaptation
\cite{mcmahan2017fedavg,niknam2020federated,lim2020federated}.
This paradigm combines the closed-loop autonomy required for ASC with
distributed knowledge sharing, allowing agents to adapt locally while
coordinating their learning at the network level. Importantly, however,
applying federated learning to interference-coupled wireless networks introduces
challenges that differ from conventional federated settings: local environments
are heterogeneous, and the experience acquired by each agent depends not only
on its own spectrum decisions but also on the decisions of neighboring agents
through their shared interference environment
\cite{li2020fedprox,du2022multi,adeogun2023distributed}.

Operating effectively under these conditions requires each agent to build its
decisions from limited local information while retaining sufficient experience
to learn the longer-term consequences of its actions. A network manager does
not observe the underlying interference field directly. Instead, its users
sense local channel conditions and report highly compressed indicators of
whether their SINR satisfies the required QoS level. The manager aggregates
these reports into a local quality representation that guides its spectrum
decisions (Section~\ref{subsec:model_objective}). This representation is
intentionally compact, but necessarily provides only a partial view of the
shared wireless environment. Moreover, the information available to different
managers is inherently heterogeneous, since their observations depend on their
local topology, interference conditions, and served users
\cite{naparstek2018deep,adeogun2022multi}.

Learning from such observations requires experience to be accumulated over
time. The consequence of a spectrum decision may extend beyond its immediate
effect: a channel change modifies the interference experienced by neighboring
networks, which may subsequently adapt their own decisions and thereby alter
the environment observed by the original agent. A single interaction therefore
provides only limited information about the longer-term value of a spectrum
decision. Rather than pooling these interactions in a centralized replay
memory, the federated setting considered in this work allows each agent to
maintain a persistent local memory of its own experience. This design keeps raw
observations, actions, and rewards at their originating networks while allowing
agents to continually refine their local decision policies
\cite{zhuo2019federated,qi2021federated}.

The resulting learning problem is particularly demanding because value
estimation must remain effective under partial observability and
interference-coupled local experience. To support stable local learning, we
employ a target-network-free Mellowmax value learner
(Section~\ref{subsec:method_drl}). Its non-expansion backup operator provides a
stable basis for extended phases of local model updates, which become
especially important when communication with the federating entity is
intentionally infrequent~\cite{kim2019deepmellow,asadi2017alternative}.

Local adaptation alone, however, cannot provide effective spectrum
coordination. The knowledge acquired independently by the agents must be
combined so that their policies capture the interference and spatial-reuse
relationships that extend beyond any single agent's observations. In the
proposed federated architecture, this coordination is achieved during training
through periodic aggregation of locally learned model parameters, rather than
through the exchange of raw sensing measurements or interaction experiences
\cite{mcmahan2017fedavg,li2020fedprox}. At execution time, each network manager
acts autonomously using only its locally available information, without
requiring message exchange with neighboring agents or a centralized controller.

This architecture creates a fundamental tension between local learning and
global coordination. Increasing the number of local updates between aggregation
rounds reduces the frequency of model exchange and allows agents to exploit
their persistent local experience, but it also increases the risk that their
models drift apart as they learn from heterogeneous and interference-coupled
local experience. Conversely, frequent aggregation can improve alignment among
agents at the cost of greater communication
\cite{konecny2016communication,kairouz2021advances,li2020fedprox}.
The central challenge is therefore not merely to federate the learning process,
but to preserve effective collective spectrum coordination when agents optimize
over heterogeneous local experience that is coupled through interference.
Addressing this challenge forms the basis for the FASCIA framework developed
in this paper.

%------------------------------------------------------------------------------
\subsection{Main Results}
\label{subsec:intro_main}
%------------------------------------------------------------------------------

Motivated by the challenges above, this work investigates how federated
agentic AI can enable effective adaptive spectrum coordination when autonomous
network managers learn from heterogeneous local experience while remaining
coupled through a shared interference environment. The objective is to retain
the benefits of collective learning without requiring centralized collection
of raw sensing observations or interaction histories, while accounting for the
partial observability, interference coupling, and communication constraints
inherent to distributed wireless operation. The main contributions and results
of this work are summarized as follows.\vspace{0.2cm}

\noindent
\textbf{1) Federated Agentic AI Framework for Interference-Coupled ASC:}
We formulate a challenging federated agentic AI setting for ASC, in which autonomous network managers learn and act based on
partial local SINR observations while interacting through a shared interference
environment. Unlike conventional federated learning settings, the agents do
not interact with isolated local environments. Their operating conditions are
heterogeneous, and the SINR observations available to each agent depend jointly
on its own spectrum decisions and those of neighboring agents. Consequently,
the local experiences used for learning are intrinsically coupled through the
shared wireless medium. At the same time, collecting raw sensing observations
and interaction histories at a central entity is undesirable due to
communication overhead and privacy concerns. The resulting framework therefore
requires agents to learn from heterogeneous and interference-coupled local
experience, benefit from collective knowledge while limiting information
exchange, and ultimately execute their spectrum decisions autonomously. To the
best of our knowledge, this is the first federated agentic AI framework to
address this challenging interference-coupled spectrum coordination setting.\vspace{0.2cm}

\noindent
\textbf{2) Algorithm Development for Coupled and Heterogeneous Learning:}
To address this setting, we develop \OURS{} (\emph{Federated Agentic Spectrum
Coordination with Interference Awareness}), a federated multi-agent DRL
algorithm tailored to the distinctive learning dynamics of
interference-coupled ASC. Each agent continuously accumulates its locally
acquired experience in a persistent replay buffer, retained across interaction
episodes, and employs a target-network-free Mellowmax value learner to derive
its spectrum-coordination policy under partial observability. At each federated
round, agents perform $E$ local learning updates using exclusively their own
accumulated experience and periodically exchange only their learned model
parameters for global aggregation; raw observations, actions, rewards, and
transition tuples remain local. This design eliminates the need for centralized
experience pooling while allowing agents to benefit from knowledge acquired
across the network.

The local learning problems, however, can differ substantially across network
topologies, loads, user populations, and interference conditions. Moreover,
the experience collected by different agents is not independent: the SINR
measurements and rewards observed by one agent depend on the spectrum decisions
of neighboring agents through inter-network interference. Consequently, prolonged local optimization over
heterogeneous and interference-coupled experience can drive
the locally learned models in different directions, resulting in
substantial model drift and degrading collective learning. To
mitigate this effect, FASCIA incorporates proximal regularization
into the federated optimization to penalize excessive deviation
of locally updated models from the global model. We further investigate the effect of the number of local updates $E$ on
coordination performance under heterogeneous and
interference-coupled local experience. The resulting algorithm supports
collective learning during training while enabling each network manager to
execute its learned spectrum policy autonomously and without inter-agent
message exchange after training.\vspace{0.2cm}

\noindent
\textbf{3) Comprehensive Performance Evaluation:}
We conduct an extensive simulation study across heterogeneous network
configurations, interference conditions, and deployment scales to evaluate both
the spectrum-coordination performance and the learning behavior of \OURS{}. The results demonstrate that FASCIA achieves performance close to the centralized
benchmark despite relying on partial local observations and without
centrally pooling raw interaction experience. We further compare alternative federated aggregation
strategies and evaluate the effect of the number of local updates
on learning behavior and coordination performance. The results show that \OURS{} maintains strong
performance across heterogeneous and interference-coupled operating conditions
and generalizes effectively to network configurations beyond those encountered
during training. Overall, the experiments demonstrate that effective collective
spectrum coordination can be achieved through federated agentic learning while
keeping raw network experience local and retaining autonomous, message-free
execution.\vspace{0.2cm}

\noindent
\textbf{4) Open-Source Implementation:}
To facilitate reproducibility and further research, we make the implementation
of \OURS{}, together with the simulation and evaluation environment, publicly
available as open-source code at \cite{baziza2026code}. The released code provides
the training and evaluation pipeline required to reproduce the reported results
and offers the research community a platform for further investigation of
federated agentic AI, multi-agent learning, and adaptive spectrum coordination
in interference-coupled wireless networks.

\subsection{Related Work}
\label{subsec:intro_related}
%------------------------------------------------------------------------------

\begin{table*}[!t]
\centering
\caption{Comparison with representative learning-based spectrum and
wireless resource coordination approaches.}
\label{tab:related}
\footnotesize
\begin{tabular}{@{}lcccccc@{}}
\toprule
\textbf{Work} &
\textbf{Distributed} &
\textbf{Federated} &
\textbf{Experience} &
\textbf{ICI-} &
\textbf{Spectrum-mobility} &
\textbf{Training} \\
&
\textbf{execution} &
\textbf{learning} &
\textbf{remains local} &
\textbf{aware} &
\textbf{awareness} &
\textbf{paradigm} \\
\midrule

QDSA~\cite{naparstek2019deep}
& Yes & No & No & No & No & Centralized DQN \\

MADDQN-DCA~\cite{adeogun2023distributed}
& Yes & No & No & No & No & Multi-agent DDQN \\

IDSA~\cite{kaur2023deep}
& Yes & No & No & No & No & Deep recurrent RL \\

MARL-RM~\cite{du2022multi}
& Yes & No & No & No & No & Multi-agent RL \\

\CARLTON{}~\cite{cohen2024carlton}
& Yes & No & No & Yes & Yes & CTDE \\

FRL-V2X~\cite{xu2024federated}
& Yes & Yes & Yes & No & No & Federated RL \\

FeDRL-D2D~\cite{noman2024fedrl}
& Yes & Yes & Yes & No & No & Federated DDQN \\

\midrule
\textbf{\OURS{} (this work)}
& \textbf{Yes}
& \textbf{Yes}
& \textbf{Yes}
& \textbf{Yes}
& \textbf{Yes}
& \textbf{Federated Mellowmax DRL} \\

\bottomrule
\end{tabular}
\end{table*}

Learning-based spectrum coordination has been extensively studied using
multi-armed bandits (MABs), restless MABs (RMABs), game-theoretic methods,
and, more recently, deep reinforcement learning (DRL). Representative MAB/RMAB
approaches include
\cite{liu2012cooperative,Tekin_2012_Online,liu2012learning,
nayyar2016regret,avner2016multi,bistritz2018distributed,
gafni2020learning,wang2021decentralized,gafni2022learning},
while distributed spectrum-learning and access mechanisms have also been
developed using stable matching and game-theoretic optimization
\cite{leshem2012multichannel,gafni2022distributed,
cohen2013game,cohen2016distributedToN,cohen2017distributed,
malachi2020queue}. DRL and multi-agent RL (MARL) subsequently enabled spectrum policies to be
learned directly from interaction
\cite{naparstek2017deep,naparstek2018deep,du2022multi,
adeogun2022multi,adeogun2023distributed,kaur2023deep,
bokobza2023deep}. In particular, \CARLTON{}~\cite{cohen2024carlton}
developed a centralized-training-with-decentralized-execution (CTDE)-based MARL framework for distributed channel allocation under
partially overlapping channels and inter-carrier interference. These works
establish the effectiveness of learning-based spectrum adaptation, but the
training of high-performance MARL policies commonly relies on centrally
collected experience or centralized learning.

Federated learning (FL) provides an alternative approach in which clients train
locally and periodically aggregate model information without centrally pooling
their raw data. \FedAvg{}~\cite{mcmahan2017fedavg} established the standard
federated averaging paradigm, while subsequent work studied communication
efficiency and learning under heterogeneous client data
\cite{konecny2016communication,kairouz2021advances}. In particular,
\FedProx{}~\cite{li2020fedprox} penalizes excessive local deviation from the
global model, while SCAFFOLD~\cite{karimireddy2020scaffold} uses control
variates to reduce client drift. FL has also been widely investigated for
wireless and edge intelligence
\cite{niknam2020federated,lim2020federated,chen2021joint,park2019wireless}.
Federated reinforcement learning extends this idea to agents that generate
their own experience through interaction
\cite{zhuo2019federated,qi2021federated}, with recent applications to wireless
resource allocation in V2X and D2D-assisted networks
\cite{xu2024federated,noman2024fedrl}. Unlike conventional FRL settings in which
clients typically learn from distinct local environments, the network managers
considered here interact through the same wireless medium: the SINR
measurements and rewards observed by one agent depend directly on the spectrum
decisions of neighboring agents. Hence, the local learning experiences are not
only heterogeneous but also intrinsically interference-coupled.

More recently, agentic AI has emerged as a framework for autonomous and
closed-loop operation of next-generation wireless networks, including
AI-native RAN control, autonomous network management, and multi-agent network
optimization
\cite{elkael2025agentrAN,ameur2026agentic}.
Of particular relevance, recent work has explicitly investigated federated
agentic AI for wireless networks~\cite{cai2026federated}, illustrating how
federated learning can support distributed agentic intelligence without
centralizing raw local data. These emerging studies, however, do not address
the interference-coupled adaptive spectrum coordination problem considered
here, where the actions of one autonomous network manager directly affect the
observations and learning experience of the others. \OURS{} addresses this
intersection through federated multi-agent DRL with persistent local experience,
model-level coordination, and explicit treatment of inter-carrier interference.

Table~\ref{tab:related} positions \OURS{} against the most closely related
representative learning-based approaches for spectrum and wireless resource
coordination. The comparison considers distributed execution, federated
learning, explicit awareness of ICI between
partially overlapping or adjacent channels, whether raw interaction experience
remains local rather than being transferred to a centralized learner, and
spectrum-mobility awareness, i.e., whether unnecessary channel changes are
explicitly considered by the learning or decision mechanism. As shown in
Table~\ref{tab:related}, existing approaches capture different subsets of these
properties. In particular, \CARLTON{} considers ICI and spectrum mobility but
relies on centralized training, whereas recent federated RL approaches enable
distributed learning with local experience but do not explicitly address ICI
between partially overlapping channels or unnecessary spectrum mobility.
To the best of our knowledge, \OURS{} is the first federated agentic AI
framework that jointly addresses these aspects for interference-coupled
adaptive spectrum coordination.

The remainder of this paper is organized as follows.
Section~\ref{sec:system} formalizes the system model and the constrained DCA
objective, and maps them onto a federated optimization problem.
Section~\ref{sec:method} presents the \OURS{} algorithm.
Section~\ref{sec:experiments} reports the experimental study, and
Section~\ref{sec:conclusion} concludes the paper.

%==============================================================================
\section{System Model and Problem Formulation}
\label{sec:system}
%==============================================================================

%------------------------------------------------------------------------------
\subsection{Description of the System}
\label{subsec:model_system}
%------------------------------------------------------------------------------

\begin{table}[!t]
\centering
\caption{Summary of the main notation.}
\label{tab:notation}
\footnotesize
\begin{tabular}{@{}ll@{}}
\toprule
\textbf{Symbol} & \textbf{Meaning} \\
\midrule
$N$, $\mathcal{N}^{n}$ &
number of networks; network $n$ \\

$M^{n}$, $M_{\max}$ &
number of users in $\mathcal{N}^{n}$; maximum number \\

$K$, $B$ &
number of channels; per-channel bandwidth \\

$c^{n}$ &
operating channel of $\mathcal{N}^{n}$ \\

$T(k,\tilde{k})$ &
ICI attenuation between channels $k$ and $\tilde{k}$ \\

$\mathrm{SINR}^{*}$ &
minimum network-level QoS-SINR requirement \\

$\mathrm{SINR}_{T}$ &
SINR threshold used to construct $\mathrm{BSINR}$ \\

$\mathrm{BSINR}^{n}_{j,k}$ &
binary QoS report of user $j$ on channel $k$ \\

$\mathrm{QV}^{n}$ &
quality vector of $\mathcal{N}^{n}$ \\

$s^{n}(\tau)$, $a^{n}(\tau)$ &
state and action of agent $n$ at step $\tau$ \\

$r_p$ & personal reward of the local agent \\
$L_{\mathrm{CBR}}$ & number of bits in the channel binary representation \\

$\mathcal{D}^{n}$, $S_{z}$ &
local replay buffer of client $n$; buffer capacity \\

$w$, $d_{w}$ &
model parameters; number of model parameters \\

$T$, $t$ &
decision steps per episode; federated round index \\

$E$, $b_{z}$ &
local update steps per round; mini-batch size \\

$\mathcal{C}_{t}$, $N_{t}$ &
participating clients in round $t$; their number \\

$F_{n}(w)$, $F(w)$ &
local and global federated objectives \\

$\mu$ &
\FedProx{} proximal coefficient \\

$\omega$ &
mellowmax temperature \\
\bottomrule
\end{tabular}
\end{table}

We consider a shared-spectrum environment consisting of $N$ coexisting
wireless networks, represented by
$\mathcal{N}=\{\mathcal{N}^{1},\ldots,\mathcal{N}^{N}\}$. Each network
$\mathcal{N}^{n}$ serves $M^{n}$ users, where
$M^{n}\in\{1,2,\ldots,M_{\max}\}$ and $M_{\max}$ is the maximum number of
users among the networks. The total available bandwidth is partitioned into
$K$ equal-sized overlapping channels of bandwidth $B$, indexed by
$\mathcal{K}=\{1,2,\ldots,K\}$. At each time slot, every network operates on
a single channel.

\begin{figure}[!t]
\centering
\resizebox{0.94\columnwidth}{!}{%
\begin{tikzpicture}[scale=1.0, font=\small]
  \def\sig{0.42}
  \def\amp{1.25}
  \draw[-{Latex[length=1.8mm]}] (-0.35,0) -- (7.9,0)
        node[below right=0pt and -6pt, font=\scriptsize] {frequency};
  \draw[-{Latex[length=1.8mm]}] (-0.35,0) -- (-0.35,2.25);
  \node[font=\scriptsize, rotate=90, anchor=south] at (-0.72,0.9) {PSD};
  \begin{scope}
    \clip (2.55,0) rectangle (3.45,2.0);
    \fill[red!25]
      plot[domain=2.5:3.5,samples=60] (\x,{\amp*exp(-((\x-2.0)/\sig)^2)})
      -- (3.5,0) -- (2.5,0) -- cycle;
  \end{scope}
  \foreach \i/\c in {0/blue,1/teal,2/violet,3/orange,4/brown,5/olive,6/purple}{
    \draw[\c!75, thick]
      plot[domain={\i-1.55}:{\i+1.55},samples=80]
      (\x,{\amp*exp(-((\x-\i)/\sig)^2)});
    \draw[gray!55, dotted] (\i,0) -- (\i,{\amp*1.02});
  }
  \foreach \i/\lab in {0/$f_1$,1/$f_2$,2/$f_3$,3/$f_4$,5/$f_{K-1}$,6/$f_K$}{
    \draw (\i,0) -- (\i,-0.09) node[below, font=\scriptsize] {\lab};
  }
  \node[font=\scriptsize] at (4,-0.3) {$\cdots$};
  \foreach \i/\lab in {0/$C_1$,1/$C_2$,2/$C_3$,3/$C_4$,6/$C_K$}{
    \node[font=\scriptsize, anchor=south] at (\i,{\amp*1.03}) {\lab};
  }
  \draw[<->, gray!75] (0,-0.62) -- (1,-0.62)
        node[midway, below, font=\scriptsize] {$f_{k+1}-f_{k}$};
  \node[font=\scriptsize, red!65!black, align=center, anchor=west]
        at (3.7,2.02) {ICI: residual overlap attenuated by $T(k,\tilde{k})$};
  \draw[->, red!65!black, thin] (3.65,1.98) .. controls (3.2,1.6) .. (3.05,0.5);
\end{tikzpicture}}
\caption{Illustration of the partially overlapping, non-orthogonal channel
partition. Channel $i$, denoted by $C_i$, corresponds to carrier frequency
$f_i$.}
\label{fig:channels}
\end{figure}
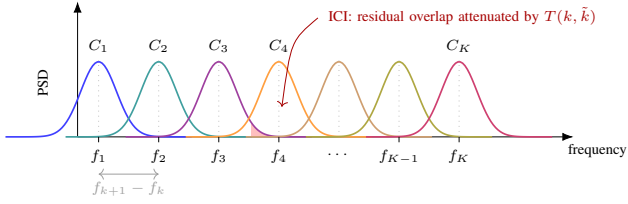

A key characteristic of the considered spectrum environment is that the
channels are \emph{non-orthogonal}. Consequently, transmissions on different
channels are not necessarily isolated from one another, and ICI may arise between partially overlapping channels. The
channels are indexed according to their serial positions in $\mathcal{K}$, as
illustrated in Fig.~\ref{fig:channels}. This setting leads to an
interference-coupled spectrum-coordination problem in which the communication
quality experienced by one network depends not only on its own channel
selection, but also on the channels occupied by the other networks.

We characterize the interference among the networks using the Egli propagation
model~\cite{egli1957radio}, a well-established model for outdoor UHF and VHF
radio propagation. Accordingly, the SINR experienced by each user on each
available channel incorporates both the aggregate interference generated by
the other networks and the additive thermal noise at the receiver.

To describe the corresponding quantities, we use the indexing notation
\begin{equation}
(\cdot)^{l,n}_{m,j,k}
\end{equation}
for a parameter associated with a pair of users
$m\in\mathcal{N}^{l}$ and $j\in\mathcal{N}^{n}$, evaluated with respect to
channel $k$. When a quantity is associated with only one user, the second user
index is omitted.

Consider users $i,j\in\mathcal{N}^{n}$, where user $i$ transmits a message to
user $j$ while network $\mathcal{N}^{n}$ operates on channel
$k\in\mathcal{K}$. The SINR measured at the receiver of user $j$ is
\begin{equation}
\text{SINR}^{n,n}_{i,j,k} =
\frac{\text{PR}^{n,n}_{i,j,k}}{I_T + I^{n}_{j,k}},
\end{equation}
where $\text{PR}^{n,n}_{i,j,k}$ denotes the received signal power,
$I_T$ is the additive thermal noise power, and $I^{n}_{j,k}$ is the
interference power observed by user $j\in\mathcal{N}^{n}$ on channel $k$.
The latter accounts for transmissions originating from all networks in
$\mathcal{N}\setminus\mathcal{N}^{n}$, including ICI caused by transmissions
on partially overlapping channels. Interference among users belonging to the
same network is assumed to be handled by the Medium Access Control (MAC)
scheduler.

The received signal power is computed as
\begin{equation}\label{PR}
\text{PR}^{n,n}_{i,j,k}= \text{PT}^{n}_{j,k} - \text{PL}^{n,n}_{i,j,k}
- \chi_{i,j,k}^{n,n} - \psi_{i,j,k}^{n,n},
\end{equation}
where $\text{PT}^{n}_{j,k}$ is the transmitter power of user
$j\in\mathcal{N}^{n}$ on channel $k$ (possibly a broadcast signal
irrespective of user $j\in\mathcal{N}^{n}$) in dBW,
$\text{PL}^{n,n}_{i,j,k}$ denotes the path loss between users
$j\in\mathcal{N}^{n}$ and $i\in\mathcal{N}^{n}$ on channel $k$, and
$\chi_{i,j,k}^{n,n}$ and $\psi_{i,j,k}^{n,n}$ represent the small-scale
fading and shadowing effects, respectively.

The interference produced by the remaining networks
$\mathcal{N}\setminus\mathcal{N}^{n}$ at user $j\in\mathcal{N}^{n}$ on
channel $k$ is
\begin{equation}
\begin{gathered}
I^{n}_{j,k} = \sum_{l \in \{1,...,n-1,n+1,...,N\}} \Phi^{n, l}_{j, k} \\
\hspace{-7.5cm}\mbox{where}\\
\Phi^{n, l}_{j, k} \triangleq
\sum_{m \in \{1,...,M^{l}\}} \phi^{n, l}_{j, m, k},
\end{gathered}
\end{equation}
where $\phi^{n,l}_{j,m,k}$ denotes the interference contribution of user
$m\in\mathcal{N}^{l}$ and is given by
\begin{equation}
\phi^{n, l}_{j, m, k} =
\text{PR}^{n, l}_{j, m, k} - T(k,\tilde{k}),
\end{equation}
where $\tilde{k}$ is the channel occupied by network $\mathcal{N}^{l}$ and
$T(k,\tilde{k})$ is the attenuation observed at channel $k$ with respect to
channel $\tilde{k}$~\cite{goldsmith2005wireless}. Hence, the effect of an interfering network is determined
not only by its received power but also by the spectral relation between its
operating channel and the channel of interest. This explicitly captures the
ICI associated with the partially overlapping channel structure.

%------------------------------------------------------------------------------
\subsection{Federated Agentic Learning Model}
\label{subsec:model_federated}
%------------------------------------------------------------------------------

Spectrum coordination is carried out by a network manager associated with each
network $\mathcal{N}^{n}$. In the proposed framework, the network manager acts
as an autonomous AI agent: it obtains measurements generated within its own
network, selects the operating channel, observes the resulting communication
quality, and uses this interaction experience to improve its subsequent
spectrum decisions. Thus, the agents operate locally while remaining coupled
through the shared wireless environment, since the channel selected by one
network affects the interference experienced by the others.

Unlike centralized-training approaches, we do not assume that the interaction
experience generated by the different network managers is collected by a
central learner. Instead, network manager $n$ is identified with federated
client $n$ and retains locally the information generated through its
interaction with the spectrum environment. In particular, its per-user
$\text{BSINR}$ measurements, locally constructed observations, selected
channels, received rewards, and stored interaction experience remain at the
corresponding network. Coordination of the learning process is performed at
the model level through periodic federated aggregation rather than by pooling
the agents' raw experience.

This architecture is particularly relevant to the interference-coupled setting
considered here. Although the agents ultimately learn a common spectrum
coordination capability, the experience available to different network
managers need not follow identical distributions. Networks may serve different
numbers of users, experience different interference neighborhoods, and observe
different channel conditions. Moreover, the actions taken by neighboring
agents directly affect the local measurements observed by a given network.
Consequently, the federated clients learn from heterogeneous and
interference-coupled local experience.

The above architecture specifies the information structure under which
learning takes place: interaction experience remains local, while model
information is periodically exchanged for federated learning. The precise
local learning objective, model aggregation procedure, and the resulting
\OURS{} algorithm are developed in Section~\ref{sec:method}. We emphasize that
retaining raw experience locally provides an architectural separation between
the participating networks; federation by itself, however, does not constitute
a formal privacy guarantee.

%------------------------------------------------------------------------------
\subsection{Problem Formulation and Objective}
\label{subsec:model_objective}
%------------------------------------------------------------------------------

Let $c^{n}\in\{1,...,K\}$ denote the operating channel selected by network
$\mathcal{N}^{n}$, and let ${\text{SINR}}^{n}_{j,c^n}$ denote the average SINR
of user $j\in\mathcal{N}^{n}$, accounting for the messages received from the
other $M^{n}-1$ users in the same network. It is given by
\begin{equation}
{\text{SINR}}^{n}_{j,c^n} =
\frac{1}{M^{n}-1}\sum_{\substack{i=1 \\ i\neq j}}^{M^{n}}
\text{SINR}^{n,n}_{j, i, c^n}.
\end{equation}
The corresponding average SINR of network $\mathcal{N}^{n}$, denoted by
$\overline{\text{SINR}}^{n}_{c^n}$, is
\begin{equation}
\overline{\text{SINR}}^{n}_{c^n} =
\frac{1}{M^{n}}\sum_{j=1}^{M^{n}}
{\text{SINR}}^{n}_{j,c^n}.
\end{equation}

At the network level, the adaptive spectrum coordination objective is to select
the operating channels so as to maximize the average communication quality
across the participating networks, while requiring every network to satisfy a
target QoS-SINR level, denoted by $\text{SINR}^{*}$. Accordingly, the desired
spectrum configuration is characterized by
\begin{equation}
\label{eq:objective}
\begin{array}{l}
\displaystyle \max_{\left\{c^{n}\right\}^N}
\frac{1}{N}\sum_{n=1}^{N}\overline{\text{SINR}}^{n}_{c^n}
\vspace{0.2cm}\\\displaystyle
\text{subject to }
\overline{\text{SINR}}^{n}_{c^n} \geq \text{SINR}^*
\ \;\;\forall n\in \mathcal{N}.
\end{array}
\end{equation}

The QoS constraints in \eqref{eq:objective} are important because unconstrained
maximization of the average SINR can favor some networks at the expense of
others. Requiring every network to exceed the prescribed QoS-SINR threshold
prevents an improvement in aggregate performance from being obtained by
allowing individual networks to experience unacceptable communication quality.
This formulation therefore combines high aggregate performance with a minimum
QoS requirement for each participating network.

We further consider a setting in which only a very small amount of measurement
information is exchanged within each network. To this end, the SINR
measurement is binarized with respect to a target QoS-SINR sensitivity
threshold $\text{SINR}_{T}$:
\begin{equation}
\label{eq:BSINR}
\text{BSINR}^{n}_{j,k} =
\mathbbm{1}({\text{SINR}}^{n}_{j,k} > \text{SINR}_T),
\end{equation}
where $\mathbbm{1}$ is the indicator function, equal to one when
${\text{SINR}}^{n}_{j,k}>\text{SINR}_{T}$ and zero otherwise.

Under this limited-information representation, the spectrum-coordination
objective becomes
\begin{equation}
\label{eq:BSINR_objective}
\begin{array}{l}
\displaystyle \max_{\left\{c^{n}\right\}^N}
\frac{1}{N}\sum_{n=1}^{N}
\overline{\text{BSINR}}^{n}_{c^n}
\vspace{0.2cm}\\\displaystyle
\text{subject to }
\overline{\text{SINR}}^{n}_{c^n} \geq \text{SINR}^*
\ \;\;\forall n\in \mathcal{N},
\end{array}
\end{equation}
where
$\overline{\text{BSINR}}^{n}_{c^n}\triangleq
\frac{1}{M^{n}}\sum_{j=1}^{M^{n}}\text{BSINR}^{n}_{j,c^{n}}$.

Equations~\eqref{eq:objective}--\eqref{eq:BSINR_objective} characterize the
desired network-level outcome. Importantly, however, we do not assume that the
channel vector $\{c^{n}\}_{n=1}^{N}$ is repeatedly computed by a centralized
optimizer. Instead, each network manager must learn how to select its operating
channel from locally available information. Hence, the learning objective is
to obtain, through federated training, a model that enables the autonomous
agents to collectively realize high-quality spectrum configurations according
to \eqref{eq:BSINR_objective}, while retaining their locally generated
interaction experience within their respective networks. This distinction
between the operational spectrum objective and the federated learning
mechanism is central to the proposed \OURS{} framework.

We assume that users can measure SINR values across all available channels.
Although wideband spectrum sensing has traditionally presented implementation
challenges, including power consumption and signal-processing requirements,
advances in processing capabilities, spectrum-sensing algorithms, and
software-defined radio (SDR) have made flexible wideband measurements
increasingly practical~\cite{arjoune2019comprehensive,nasser2021spectrum}.

Finally, the considered setting places two important requirements on the
learning framework. First, spectrum decisions should adapt rapidly to changes
in the interference environment, motivating a short learning and adaptation
time. Second, channel selection must be executable locally by the network
managers to support scalable operation without relying on a centralized
decision maker. These requirements, together with the federated information
structure described above, motivate the learning and coordination mechanism
developed in the following sections.

%==============================================================================
\section{The Federated Agentic Spectrum Coordination
with Interference Awareness (\OURS{}) Algorithm}
\label{sec:method}
%==============================================================================

%------------------------------------------------------------------------------
% FIGURE: CTDE vs FRL
%------------------------------------------------------------------------------
\begin{figure*}[!t]
\centering
\resizebox{0.92\textwidth}{!}{%
\begin{tikzpicture}[
  font=\small,
  box/.style   = {rounded corners=2pt, draw=black!65, fill=white,
                  minimum width=20mm, minimum height=8mm, align=center,
                  inner sep=3pt},
  agent/.style = {box, fill=blue!8, draw=blue!55},
  serverc/.style= {box, fill=red!8, draw=red!60, minimum width=36mm},
  serverf/.style= {box, fill=green!8, draw=green!55!black, minimum width=36mm},
  bufc/.style  = {box, fill=red!6, draw=red!55, font=\scriptsize,
                  minimum width=20mm, minimum height=7mm},
  buff/.style  = {box, fill=orange!10, draw=orange!70!black, font=\scriptsize,
                  minimum width=20mm, minimum height=7mm},
  raw/.style   = {-{Latex[length=2mm]}, draw=red!70, thick},
  wgt/.style   = {-{Latex[length=2mm]}, draw=green!45!black, thick},
  bcast/.style = {-{Latex[length=2mm]}, draw=blue!65, thick, dashed},
  loc/.style   = {{Latex[length=1.6mm]}-{Latex[length=1.6mm]}, draw=black!55},
  lbl/.style   = {font=\scriptsize, inner sep=1pt}
]

%%%%%%%%%%%%%%%%%%%%%%% (a) CENTRALIZED CTDE %%%%%%%%%%%%%%%%%%%%%%%%%%%%%%%%%
\begin{scope}
  \node[serverc] (grm) at (0,0) {Global Replay Memory};
  \node[serverc, below=6mm of grm, fill=red!14] (learner)
        {Central Learner\\[-1pt]{\scriptsize all $N_E$ gradient updates}};

  \node[agent] (a1) at (-3.3,3.0) {NM $1$};
  \node[agent] (a2) at ( 0.0,3.0) {NM $2$};
  \node[agent] (a3) at ( 3.3,3.0) {NM $N$};
  \node[lbl]   at (1.65,3.0) {$\cdots$};

  \node[bufc, above=4mm of a1] (d1) {Temporary buffer $\mathcal{D}^{1}$};
  \node[bufc, above=4mm of a2] (d2) {Temporary buffer $\mathcal{D}^{2}$};
  \node[bufc, above=4mm of a3] (d3) {Temporary buffer $\mathcal{D}^{N}$};
  \draw[loc] (a1) -- (d1);  \draw[loc] (a2) -- (d2);  \draw[loc] (a3) -- (d3);
  \node[lbl, red!70!black, above=1mm of d2] {Local experience transferred to global memory};

  \draw[raw] (a1.south) -- node[lbl,left=1pt,pos=0.5]
        {\textcolor{red!75!black}{interaction tuple $(s,a,r,s')$}} (grm.north west);
  \draw[raw] (a2.south) -- (grm.north);
  \draw[raw] (a3.south) -- (grm.north east);

  \draw[loc, thick] (grm.south) -- node[lbl,right=1pt]
        {\scriptsize mini-batches $b_z$} (learner.north);

  \draw[bcast] (learner.west) -- ++(-2.2,0) |- (a1.west);
  \draw[bcast] (learner.east) -- ++( 2.2,0) |- (a3.east);
  \node[lbl, blue!65, below left=1mm and -3mm of a1.south west]
        {broadcast $w$};

  \node[font=\bfseries] at (0,-2.7) {(a) Centralized Training (CTDE)};
  \node[lbl, text width=68mm, align=center, red!70!black] at (0,-3.4)
        {Experience centrally collected~$\bullet$~centralized model updates~$\bullet$~global replay memory};
\end{scope}

%%%%%%%%%%%%%%%%%%%%%%% (b) FEDERATED FRL %%%%%%%%%%%%%%%%%%%%%%%%%%%%%%%%%%%%
\begin{scope}[xshift=108mm]
  \node[serverf] (agg) at (0,0)
        {Aggregation Server\\[-1pt]{\scriptsize \FedAvg{} / \FedProx{}}};
  \node[below=1.5mm of agg, font=\scriptsize, align=center]
        {$w^{t+1}=\sum_{k}\frac{n_k}{n}\,w_k^{t+1}$};

  \node[agent] (b1) at (-3.3,3.0) {NM $1$\\AI agent};
  \node[agent] (b2) at ( 0.0,3.0) {NM $2$\\AI agent};
  \node[agent] (b3) at ( 3.3,3.0) {NM $N$\\AI agent};
  \node[lbl]   at (1.65,3.0) {$\cdots$};

  \node[buff, above=4mm of b1] (p1) {Local replay buffer $\mathcal{D}^{1}$};
  \node[buff, above=4mm of b2] (p2) {Local replay buffer $\mathcal{D}^{2}$};
  \node[buff, above=4mm of b3] (p3) {Local replay buffer $\mathcal{D}^{N}$};
  \draw[loc, draw=orange!75!black] (b1) -- (p1);
  \draw[loc, draw=orange!75!black] (b2) -- (p2);
  \draw[loc, draw=orange!75!black] (b3) -- (p3);
  \node[lbl, orange!70!black, above=1mm of p2]
        {$E$ local updates -- experience remains local};

  \draw[wgt] (b1.south) -- node[lbl,left=1pt,pos=0.5]
        {\textcolor{green!35!black}{$w_1^{t+1}$}} (agg.north west);
  \draw[wgt] (b2.south) -- (agg.north);
  \draw[wgt] (b3.south) -- (agg.north east);

  \draw[bcast] (agg.west) -- ++(-2.2,0) |- (b1.west);
  \draw[bcast] (agg.east) -- ++( 2.2,0) |- (b3.east);
  \node[lbl, blue!65, below left=1mm and -3mm of b1.south west]
        {broadcast $w^{t}$};

  \node[font=\bfseries] at (0,-2.7) {(b) FASCIA: Federated Training};
  \node[lbl, text width=68mm, align=center, green!35!black] at (0,-3.4)
        {Experience remains local~$\bullet$~distributed model updates~$\bullet$~federated model aggregation};
\end{scope}

\end{tikzpicture}}
\caption{Comparison of centralized and federated training architectures for
adaptive spectrum coordination. (a)~Under CTDE, locally generated interaction experience is
collected in a global replay memory and used for centralized model updates. (b)~In \OURS{}, each network manager (NM) acts as an AI agent, maintains a
local replay buffer, performs local learning updates, and periodically
exchanges model parameters with the aggregation server. Thus, interaction
experience remains local in \OURS{}, while spectrum decisions are executed
locally by the NMs in both architectures.}
\label{fig:paradigms}
\end{figure*}
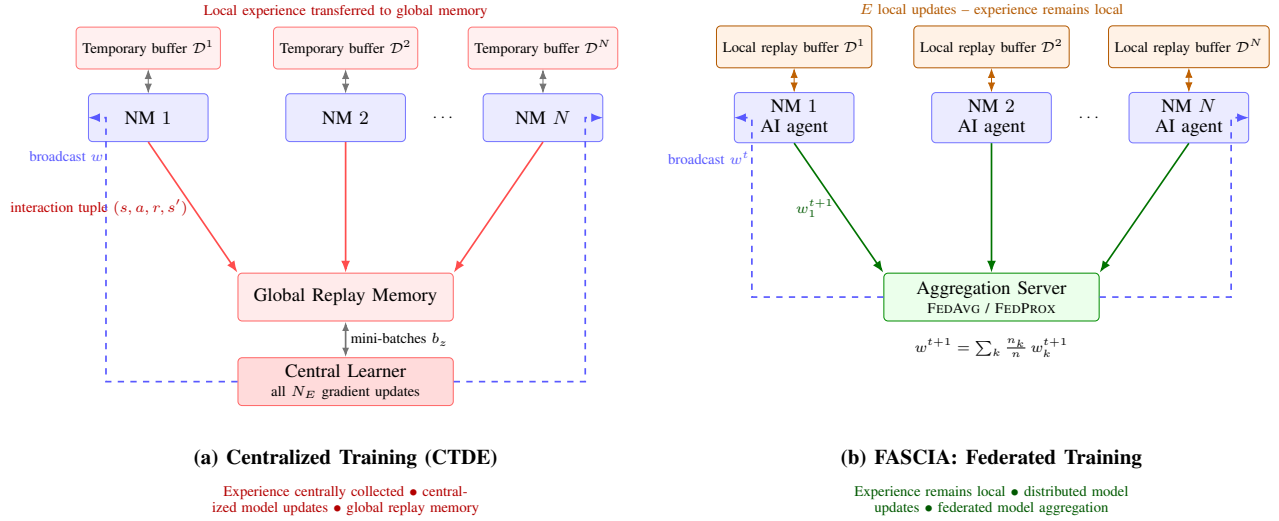

This section develops \OURS{}, a federated agentic learning framework for
interference-aware adaptive spectrum coordination. In \OURS{}, each network
manager acts as an autonomous learning agent that interacts with its local
wireless environment, maintains its own experience, and performs local
value-function updates. The agents periodically coordinate through federated
model aggregation, such that locally generated interaction experience is not
pooled at a central learner. The resulting architecture combines local
spectrum decision making with model-level coordination across the participating
networks.

Fig.~\ref{fig:paradigms} illustrates the main distinction between
the considered training paradigms. In the centralized-training architecture, the interaction tuples generated
by the network managers are collected in a global replay memory, from which a
central learner performs the model updates. In contrast, under \OURS{}, each network manager maintains a
persistent local replay buffer and performs $E$ local learning steps. Only
model parameters are periodically communicated to the aggregation server,
which combines the local models and redistributes the resulting global model.
Thus, the interaction experience used for learning remains at the network
managers throughout training.

The remainder of this section develops the components of
FASCIA. We first present the Mellowmax-based local value-learning
mechanism, followed by the agent observation and action spaces,
exploration strategy, and local reward design. We then formulate
the federated training and model-aggregation procedures and discuss
their operation under heterogeneous local experience.

\subsection{Mellowmax-Based Local Value Learning}
\label{subsec:method_drl}
%------------------------------------------------------------------------------

Each network manager operates as an autonomous AI agent that interacts with
its local wireless environment to perform adaptive spectrum coordination.
At each decision step, the agent observes its local state, selects an
operating channel, and receives a scalar reward reflecting the resulting
spectrum configuration. We adopt a value-based reinforcement-learning
approach, in which the agent estimates the action-value function $Q_{\pi}(s,a)
=
\mathbb{E}_{\pi}\!\left[
\sum_{i\geq0}\gamma^{i} r(t{+}i{+}1)
\,\middle|\,
S(t)=s,A(t)=a
\right]$, where $\gamma\in(0,1)$ is the discount factor. The action-value function is
approximated by a neural network $Q(s,a;w)$ with parameter vector $w$, whose
parameters are learned by minimizing the temporal-difference (TD)
error~\cite{sutton2018reinforcement}.

For local value learning, \OURS{} employs
DeepMellow~\cite{kim2019deepmellow}. In contrast to the hard maximization
commonly used in deep $Q$-learning, DeepMellow employs the mellowmax
operator~\cite{asadi2017alternative},
\begin{equation}
\mathrm{mm}_{\omega}\bigl(Q(s,\cdot\,;w)\bigr)
 = \frac{1}{\omega}\,
   \log\!\left(\frac{1}{K}\sum_{k=1}^{K} e^{\,\omega\, Q(s,k;w)}\right),
\label{eq:mellowmax}
\end{equation}
where $\omega$ controls the degree of smoothing. The operator approaches the
mean as $\omega\to 0$ and the maximum as $\omega\to\infty$. Moreover,
mellowmax is a non-expansion and provides a smooth alternative to the hard
maximum used in conventional $Q$-learning. Importantly for the proposed
federated architecture, DeepMellow enables local value learning using a
single parameter vector $w$, without maintaining a separate delayed target
network. Consequently, the model state periodically coordinated among the
federated agents can be represented by a single value-network parameter
vector.

Accordingly, the TD target used by client $n$ is
\begin{equation}
Y^{n}(t) = r^{n}(t+1)
  + \gamma\,\bigl(1 - d(t+1)\bigr)\,
    \mathrm{mm}_{\omega}\bigl(\mathrm{sg}\bigl(Q(s^{n}(t+1),\cdot\,;w)\bigr)\bigr),
\label{eq:target}
\end{equation}
where $\mathrm{sg}(\cdot)$ denotes the stop-gradient operator and $d$ is the
terminal indicator. Defining the TD error as
$e^{n}(t)=Y^{n}(t)-Q(s^{n}(t),a^{n}(t);w)$, we use the Huber loss
$L_{\delta=1.0}(e)$ as the local loss function. Its linear behavior for large
errors reduces the sensitivity of local updates to occasional large TD
errors, which may arise from the heterogeneous and interference-coupled
experience observed by the agents.

For the implementation evaluated in this paper, the action-value function is approximated by a three-layer residual multilayer perceptron with $128$ units per layer, Leaky-ReLU activations with slope
$0.2$, an identity output layer of width $K$, and \texttt{PyTorch}'s default
Kaiming-uniform initialization~\cite{he2015delving}. 
With input dimension $L_{\mathrm{CBR}}+K$, where
$L_{\mathrm{CBR}}=\lfloor\log_2 K\rfloor+1$ as in~\eqref{eq:state},
the number of trainable parameters is
\begin{equation}
\label{eq:dw}
d_w(K)
=
33,152 + 128L_{\mathrm{CBR}} + 257K.
\end{equation}
The same value-network architecture
is used by all learning-based methods compared in
Section~\ref{sec:experiments}.

%------------------------------------------------------------------------------
\subsection{Observation Space, Action Space, and Exploration}
\label{subsec:method_state}
%------------------------------------------------------------------------------

The local observation of each AI agent is constructed from the communication
quality experienced by its network. Specifically, using the binary SINR
indicators defined in Section~\ref{subsec:model_objective}, network manager $n$
constructs the quality vector $\mathrm{QV}^{n}\in[0,1]^K$, whose $k$th entry is
\begin{equation}
\mathrm{QV}^{n}_{k}
=
\frac{1}{M^{n}}\sum_{j=1}^{M^{n}}\mathrm{BSINR}^{n}_{j,k}.
\label{eq:qv}
\end{equation}
Thus, $\mathrm{QV}^{n}_{k}$ represents the fraction of users in network
$\mathcal{N}^{n}$ that satisfy the prescribed SINR threshold on channel $k$.
This normalized representation provides the agent with a local view of the
available spectrum quality while removing the direct dependence of the
observation dimension on the number of users in the network.

To support adaptive spectrum coordination while accounting for unnecessary
channel changes, the agent combines the quality vector with a binary
representation of its currently occupied channel. The resulting state is
\begin{equation}
\begin{array}{l}
s^{n}(t) = \bigl[\mathrm{CBR}^{n}(t); \mathrm{QV}^{n}(t)\bigr]
  \in \mathbb{R}^{\,L_{\mathrm{CBR}}+K},\;\;\\
  \hspace{4cm}
L_{\mathrm{CBR}} = \lfloor \log_{2}K \rfloor + 1 ,
\end{array}
\label{eq:state}
\end{equation}
where $\mathrm{CBR}^{n}(t)\in\{0,1\}^{L_{\mathrm{CBR}}}$ denotes the
channel binary representation of the channel currently occupied by network
$\mathcal{N}^{n}$, representing the previous action of the agent
in binary representation. Including this information enables the agent to
distinguish between retaining its current channel and switching to another
one, thereby allowing spectrum mobility to be incorporated into the learned
decision policy. The action space is the channel set $\mathcal{K}$, with
$a^{n}(t)=k$ indicating the selection of channel $k$. Accordingly, the
value network produces one action value for each of the $K$ available
channels.

Action selection combines feasibility-based masking with exploration. First,
a channel for which $\mathrm{QV}^{n}_{k}=0$ is excluded from the candidate
actions by defining
$\widetilde{Q}(s^{n},k;w) = -\infty$ if $\mathrm{QV}^{\,n}_{k}=0$ and
$Q(s^{n},k;w)$ otherwise. Hence, the agent does not explore channels for which
none of its users currently satisfies the prescribed QoS threshold. In the
degenerate case $\mathrm{QV}^{n}=\mathbf{0}$, no new channel is selected and
the current channel is retained.

Exploration over the remaining candidate channels follows a Boltzmann policy
based on the masked action values,
\begin{equation}
\Pr\bigl(a^{n}(t)=k \mid s^{n}(t)\bigr)
 = \frac{e^{\,\beta \widetilde{Q}(s^{n}(t),k;w)}}
        {\sum_{k'\in\mathcal{K}} e^{\,\beta \widetilde{Q}(s^{n}(t),k';w)}},
\label{eq:explore}
\end{equation}
which is sampled with probability $\epsilon_{b}$; otherwise, the agent selects
the greedy action. In the implementation considered in this paper,
$\epsilon_{b}$ is annealed linearly from $0.5$ to $0.01$ over the first half
of training and is held fixed thereafter. The exploration schedule is indexed
by the communication round and shared by all participating clients, such that
the agents follow the same exploration rate in each federated round while
collecting experience from their respective local environments.

%------------------------------------------------------------------------------
\subsection{Reward Design and the Locality of the Learning Signal}
\label{subsec:method_reward}
%------------------------------------------------------------------------------

The local reward guides each AI agent toward two complementary objectives of
adaptive spectrum coordination: selecting a channel that provides satisfactory
communication quality to its users and avoiding unnecessary channel changes
once a suitable spectrum configuration has been reached. To this end, \OURS{}
uses the personal reward $r_{p}$ defined in Algorithm~\ref{alg:rp}, which
depends only on the locally constructed quality vector and the channel
selections of the corresponding network manager.

For the channel selected by network $\mathcal{N}^{n}$, let
$v=\mathrm{QV}^{\,n}_{a^{n}(t)}(t)$ denote the fraction of its users satisfying
the prescribed SINR threshold. If $v\geq\zeta$, the selected channel is
considered satisfactory and the agent receives the fixed reward
$r_{\mathrm{desired}}$. Otherwise, the reward reflects the rank of the selected
channel among the $K$ available channels, mapped to $[-1,1]$. This provides a
relative learning signal that encourages the agent to favor higher-quality
channels even when none of the available channels reaches the desired quality
level.

To control spectrum mobility, an additional multiplicative bonus $c_{1}>1$ is
applied when the agent retains its previously occupied channel and the resulting
personal reward is positive. Thus, once an agent operates on a channel yielding
a favorable local outcome, remaining on that channel is preferred over an
unnecessary switch with otherwise comparable reward. The resulting mechanism
allows the learned spectrum-coordination policy to account jointly for
communication quality and channel stability.

\begin{algorithm}[!t]
\caption{Personal reward $r_{p}$ of network $\mathcal{N}^{n}$}
\label{alg:rp}
\begin{algorithmic}[1]
\Require $\mathrm{QV}^{n}(t)$, actions $a^{n}(t)$, $a^{n}(t-1)$
\Ensure  $r_{p}$
\State $v \gets \mathrm{QV}^{\,n}_{a^{n}(t)}(t)$
\If{$v \geq \zeta$}
  \State $r_{p} \gets r_{\mathrm{desired}}$
\Else
  \State $\mathrm{SQV}^{n} \gets \mathrm{sort}\bigl(\mathrm{QV}^{n}(t)\bigr)$;\;
         $i \gets 0$
  \While{$i \leq K$ \textbf{and} $v \geq \mathrm{SQV}^{n}_{i}$}
    \State $i \gets i+1$
  \EndWhile
  \State $r_{p} \gets 2\left(i/K - 0.5\right)$
       \Comment{rank of the chosen channel}
\EndIf
\If{$a^{n}(t) = a^{n}(t-1)$ \textbf{and} $r_{p} > 0$}
  \State $r_{p} \gets c_{1}\, r_{p}$ \Comment{stability bonus}
\EndIf
\end{algorithmic}
\end{algorithm}

An important property of this reward construction is that the learning signal
is entirely local. Network manager $n$ evaluates $r_{p}$ using its own quality
vector and channel history and therefore does not require the rewards,
observations, or actions of other network managers. This property is
particularly suited to the federated agentic architecture of \OURS{}, in which
the interaction experience generated by each network manager remains local.
Although the agents are coupled through the interference environment, no
exchange of their local interaction experience is required to construct the
reward. Coordination of the learning process is instead performed at the model
level through the federated training procedure developed in the following
subsections.

For the implementation evaluated in this paper, we set
$\zeta=0.9$, $r_{\mathrm{desired}}=4$, and $c_{1}=1.1$. With $T=20$
decision points per episode, the maximum attainable accumulated reward is
$T\,c_{1}r_{\mathrm{desired}}=88$, which is used in
Section~\ref{sec:experiments} as an absolute performance reference.

%------------------------------------------------------------------------------
\subsection{Federated Training Procedure}
\label{subsec:method_loop}
%------------------------------------------------------------------------------

The local learning mechanisms described above are integrated into the
federated training procedure summarized in Algorithm~\ref{alg:frl}. At
communication round $t$, the aggregation server broadcasts the current global
model $w^{t}$ to the participating network managers. Each participating
network manager then interacts with its local wireless environment as an
autonomous AI agent, performs its spectrum decisions, and stores the resulting
transitions in its persistent local replay buffer $\mathcal{D}^{n}$. For a
scenario containing $N_{t}$ networks, the interaction phase consists of
$T\cdot N_{t}$ steps, providing each agent with $T$ decision points.

Following the interaction phase, each client performs $E$ local gradient
updates using mini-batches of size $b_{z}$ sampled exclusively from its own
replay buffer $\mathcal{D}^{n}$. If the buffer contains fewer than $b_{z}$
transitions, the corresponding client skips the local update and returns the
broadcast model unchanged. After completing the local updates, client $n$
uploads its resulting model $w^{t,E}_{n}$ to the aggregation server. The
server then combines the uploaded models to obtain $w^{t+1}$, which initializes
the next communication round. Thus, spectrum interaction and experience
storage remain local to the network managers, whereas coordination across
agents is performed through periodic model aggregation.

Unlike conventional federated learning over fixed local datasets, the local
experience used by \OURS{} is generated through interaction with the wireless
environment. Consequently, the contents of $\mathcal{D}^{n}$ evolve throughout
training as the agent's spectrum-coordination policy and the interference
conditions it encounters change. Retaining the replay buffer across
communication rounds allows local updates to draw from experience accumulated
over multiple interactions rather than only from the most recently generated
episode. Moreover, the network realization is regenerated across rounds,
exposing the shared model to heterogeneous interference configurations during
federated training rather than to a single fixed deployment geometry.

The local learning objective at client $n$ is defined as the expected TD loss
over its replay buffer,
\[
F_{n}(w) =
\mathbb{E}_{(s,a,r,s')\sim\mathcal{D}^{n}}
\left[L_{\delta}\bigl(e(s,a,r,s';w)\bigr)\right].
\]
Under \FedAvg{}, each participating client performs $E$ stochastic-gradient
updates to this local objective before its model is aggregated at the server.
Because the agents collect heterogeneous and interference-coupled local
experience, however, their local optimization directions can differ
substantially. Increasing the number of local updates may therefore increase
the deviation of locally updated models from the common model $w^{t}$ before
aggregation, commonly referred to as client drift.

To control this deviation, \OURS{} employs \FedProx{}~\cite{li2020fedprox},
which augments the local objective with a proximal regularization term,
\begin{equation}
h_{n}\bigl(w; w^{t}\bigr) = F_{n}(w)
  + \frac{\mu}{2}\,\bigl\lVert w - w^{t} \bigr\rVert_{2}^{2},
\label{eq:prox_local}
\end{equation}
where $\mu\geq0$ is the proximal coefficient. The corresponding gradient
contains the additional term $\mu(w-w^{t})$, which penalizes excessive
deviation of the locally updated model from the global model received at the
beginning of the communication round. This regularization is particularly
relevant when the participating agents learn from heterogeneous local
experience, since it moderates client-specific model updates before
aggregation. Setting $\mu=0$ recovers the \FedAvg{} local objective, whereas
larger values of $\mu$ place progressively greater emphasis on remaining close
to $w^{t}$.

The resulting procedure therefore combines autonomous local interaction and
value learning with periodic model-level coordination. Each AI agent adapts
its spectrum decisions using experience generated within its own network,
while the proximal local objective limits excessive model drift during the
local-training phase. The aggregation step then integrates these locally
learned models into the global model used in the subsequent federated round.
The precise aggregation rule and its operation under heterogeneous local
experience are developed in the following subsection.

%------------------------------------------------------------------------------
\subsection{Model Aggregation under Heterogeneous Local Experience}
\label{subsec:method_agg}
%------------------------------------------------------------------------------

At the end of communication round $t$, the aggregation server receives the
locally updated model $w^{t,E}_{n}$ from each participating client
$n\in\mathcal{C}_{t}$. In the considered training procedure, every
participating AI agent contributes exactly $T$ new interaction transitions
during the round. Accordingly, the local models are combined using uniform
averaging,
\begin{equation}
w^{t+1}
=
\frac{1}{|\mathcal{C}_{t}|}
\sum_{n\in\mathcal{C}_{t}} w^{t,E}_{n}.
\label{eq:aggregation}
\end{equation}
The aggregation is applied to the complete parameter vector and does not
require the server to access the local replay buffers, observations, actions,
or rewards of the participating network managers.

Since all participating clients initialize their local optimization from the
same global model $w^{t}$, define the local model displacement as
\begin{equation}
\Delta_{n}^{t}=w^{t,E}_{n}-w^{t}.
\label{eq:local_displacement}
\end{equation}
Equation~\eqref{eq:aggregation} can then equivalently be written as
\begin{equation}
w^{t+1}
=
w^{t}
+
\frac{1}{|\mathcal{C}_{t}|}
\sum_{n\in\mathcal{C}_{t}}\Delta_{n}^{t}.
\label{eq:aggregation_displacement}
\end{equation}
This representation highlights the role of federated coordination in
\OURS{}: each network manager adapts the shared value model using its own
locally generated spectrum-interaction experience, while the server combines
the resulting model updates into a common model for the subsequent
communication round. Because the clients may experience different network
sizes, interference conditions, and spectrum configurations, their local
updates need not be identical. The proximal regularization in
\eqref{eq:prox_local} moderates excessive client-specific deviation during
local optimization, while \eqref{eq:aggregation} integrates the resulting
updates at the model level.

The complete federated training procedure is summarized in
Algorithm~\ref{alg:frl}. It combines local spectrum interaction, persistent
experience replay, Mellowmax-based value learning, proximal local updates, and
periodic model aggregation. Importantly, the interaction tuples generated by
the network managers remain in their respective local replay buffers
throughout this procedure; only the learned model parameters are exchanged
during federated coordination.

After federated training, the learned model $w^{*}$ is distributed to the network managers for decentralized execution. Each AI agent independently constructs its local state from the measurements available within its network,
applies the action mask, and evaluates the learned action-value function to select its operating channel. Hence, \OURS{} employs federated model-level coordination during training while retaining locally generated interaction
experience at the corresponding network managers. Once training is completed, the resulting ASC decisions are executed locally by the agents without requiring coordination with the aggregation server.

\begin{algorithm}[!t]
\caption{\OURS{} training procedure}
\label{alg:frl}
\begin{algorithmic}[1]
\Require Communication rounds $R$, local training steps $E$, proximal weight
         $\mu$, batch size $b_{z}$, buffer capacity $S_{z}$, schedule
         $\{\epsilon_{b}(t)\}$
\Ensure  Global model parameters $w^{*}$
\State Server initializes $w^{0}$; each network manager $n$ initializes an empty
       persistent $\mathcal{D}^{n}$ of capacity $S_{z}$
\For{$t = 0,\dots,R-1$}
  \State Sample $N_{t}$ with $N_{t} < K$, generate a scenario,
         and sample participating clients $\mathcal{C}_{t}=\{1,\ldots,N_{t}\}$
  \State \textbf{Broadcast:} $w^{t,0}_{n} \gets w^{t}$,\;
         $\forall n \in \mathcal{C}_{t}$
  \For{$\tau = 1,\dots,T$}\Comment{interaction phase}
    \For{$n \in \mathcal{C}_{t}$ \textbf{in turn order}}
      \State Collect BSINR reports; form $\mathrm{QV}^{n}$ by \eqref{eq:qv} and
             $s^{n}(\tau)$ by \eqref{eq:state}
      \State Sample $a^{n}(\tau)$ by \eqref{eq:explore} with $\epsilon_{b}(t)$;
             observe $r^{n}$, $s^{n}(\tau{+}1)$
      \State $\mathcal{D}^{n} \gets \mathcal{D}^{n} \cup
             \{(s^{n}(\tau),a^{n}(\tau),r^{n},s^{n}(\tau{+}1))\}$
    \EndFor
  \EndFor
  \For{$n \in \mathcal{C}_{t}$}\Comment{local update, in parallel}
    \If{$|\mathcal{D}^{n}| \geq b_{z}$}
      \For{$e = 1,\dots,E$}
        \State Sample a mini-batch of size $b_{z}$ from $\mathcal{D}^{n}$
        \State $w^{t,e}_{n} \gets w^{t,e-1}_{n} - \eta\,\nabla
               \bigl[\widehat{F}_{n} + \tfrac{\mu}{2}
               \lVert w^{t,e-1}_{n} - w^{t}\rVert^{2}\bigr]$
      \EndFor
    \Else
      \State $w^{t,E}_{n} \gets w^{t}$
    \EndIf
    \State Upload $w^{t,E}_{n}$
  \EndFor
  \State \textbf{Aggregate:} $w^{t+1} \gets
         \frac{1}{|\mathcal{C}_{t}|}\sum_{n \in \mathcal{C}_{t}} w^{t,E}_{n}$
\EndFor
\State \Return $w^{*} \gets w^{R}$
\end{algorithmic}
\end{algorithm}

%==============================================================================
\section{Experiments and Discussion}
\label{sec:experiments}
%==============================================================================

This section evaluates \OURS{} through extensive simulations. We first describe the simulation environment and evaluation metrics, and then examine the learning behavior and convergence of the proposed framework. Finally, we evaluate its ASC performance under different interference conditions and compare it with state-of-the-art spectrum-coordination methods. The complete
implementation of \OURS{}, together with the code required to reproduce the
experiments, is publicly available in~\cite{baziza2026code}.

%------------------------------------------------------------------------------
\subsection{Simulation Setup and Performance Metrics}
\label{subsec:exp_setup}
%------------------------------------------------------------------------------

The simulation environment follows Section~\ref{subsec:model_system}. The
available spectrum consists of $K=10$ partially overlapping channels. We use the Egli propagation model, the spectral attenuation characteristics specified in Section~\ref{subsec:model_system}, and a thermal noise floor of $I_{T}=-104.9$~dBm. In each scenario, the number of networks $N$ is drawn randomly, and network $\mathcal{N}^{n}$ serves $M^{n}\sim U\{1,\dots,15\}$ users distributed around its center according to a bivariate Gaussian distribution with covariance $\mathrm{diag}(50^{2},50^{2})~\mathrm{m}^{2}$. Network centers are placed
sequentially at a random radius $r\sim U[50,500]$~m and angle $\theta\sim U[0,2\pi]$ from a previously placed center. The network manager is the user minimizing the total distance to the other users in its network.
Each network receives $T=20$ decision points, resulting in a scenario horizon of $T\cdot N$ steps.

Training is restricted to $N\leq7$ networks, as
in~\cite{cohen2024carlton}; configurations with $N>7$ are therefore unseen during training and are used to evaluate generalization to denser deployments. Each run comprises $R=1000$ communication rounds for \OURS{} (and the
corresponding number of episodes for the centralized baseline), followed by $100$ inference episodes with fixed model parameters. The exploration probability $\epsilon_b$ is annealed from $0.5$ to $0.01$ over the first half
of training.

We evaluate two configurations of \OURS{} against the centralized CTDE reference. FASCIA-FA uses \FedAvg{} with $E=10$ local training steps and $\mu=0$, whereas FASCIA-FP uses \FedProx{} with $E=10$ and $\mu=0.01$. The selected values are examined further in
Section~\ref{subsec:exp_reward}. 

Let $c^{n}(T)$ denote the final operating channel of network
$\mathcal{N}^{n}$. The average channel quality across the participating
networks and the quality of the worst-performing network are defined as
\begin{equation}
\overline{\mathrm{CQ}}
=
\frac{1}{N}\sum_{n=1}^{N}
\mathrm{QV}^{n}_{c^{n}(T)}(T),
\qquad
\min_{\mathrm{CQ}}
=
\min_{n}
\mathrm{QV}^{n}_{c^{n}(T)}(T).
\label{eq:cq}
\end{equation}
The composite channel-quality metric is
\begin{equation}
\mathrm{CQ}_{\mathrm{comp}}
=
\mathbb{E}\!\left[
\frac{\overline{\mathrm{CQ}}+\min_{\mathrm{CQ}}}{2}
\right],
\label{eq:cq_composite}
\end{equation}
which accounts for both the average allocation quality and the performance of
the worst-served network.

The achieved communication rate is evaluated using the Shannon throughput.
For user $j$ in network $\mathcal{N}^{n}$ operating on channel $c^{n}$, the
achievable rate is
\begin{equation}
R^{n}_{j}
=
B\log_{2}\!\left(1+\mathrm{SINR}^{n}_{j,c^{n}}\right),
\label{eq:shannon_rate}
\end{equation}
where $B$ is the channel bandwidth. The reported throughput is averaged over
the users, participating networks, and evaluation scenarios.

Finally, the trained models are evaluated
for $N\in\{2,\dots,15\}$ using $30$ independent games for each network size,
for a total of $420$ evaluation games per method.

%------------------------------------------------------------------------------
\subsection{Training Dynamics and Parameter Selection}
\label{subsec:exp_reward}
%------------------------------------------------------------------------------

\begin{figure}[!t]
\centering
\includegraphics[width=\columnwidth]{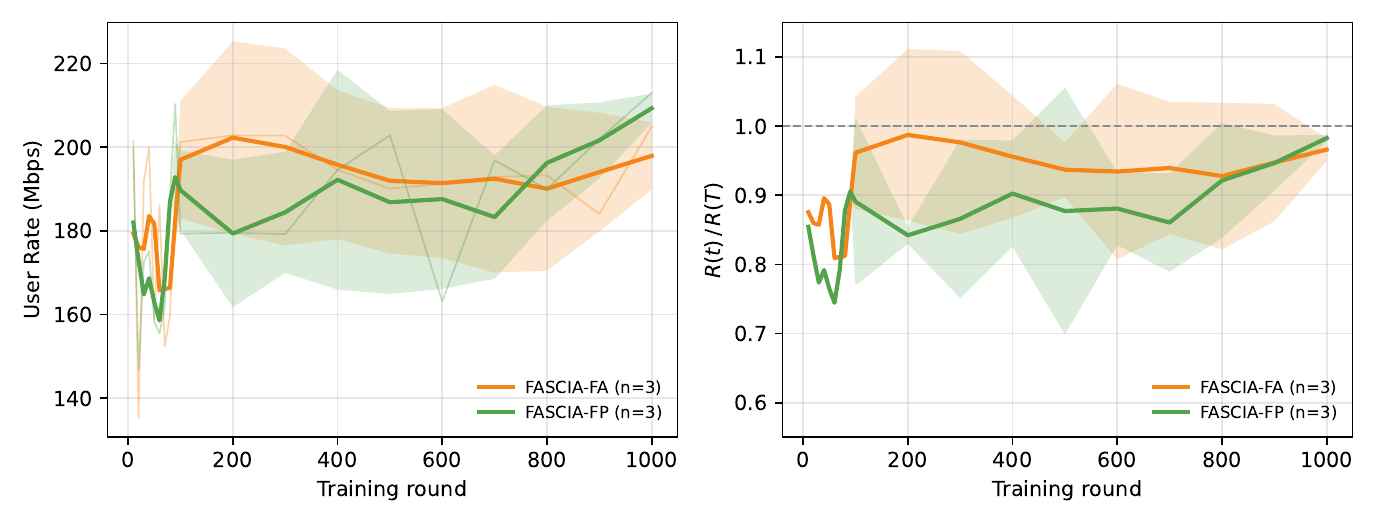}
\caption{Achievable user rate of the learned allocation.
(a)~Absolute user rate in Mbps.
(b)~Normalized learning curve $R(t)/R(T)$, where $T$ is the final training
round. Mean $\pm$ std over $3$ seeds.}
\label{fig:throughput}
\end{figure}

We first examine the evolution of the achieved Shannon throughput during
federated training. For each saved global-model checkpoint, we fix the model
parameters and evaluate the resulting spectrum allocation over a common set of
in-sample scenarios. Fig.~\ref{fig:throughput} reports the network-average
throughput as a function of the communication round for \FedAvg{} and
\FedProx{} (mean $\pm$ standard deviation over three independent runs).
Both methods improve rapidly during the first few hundred rounds and then
approach their respective operating regions. \FedAvg{} subsequently exhibits
relatively little change, whereas \FedProx{} shows a modest improvement during
the later training rounds.

\begin{figure}[!t]
\centering
\includegraphics[width=\columnwidth]{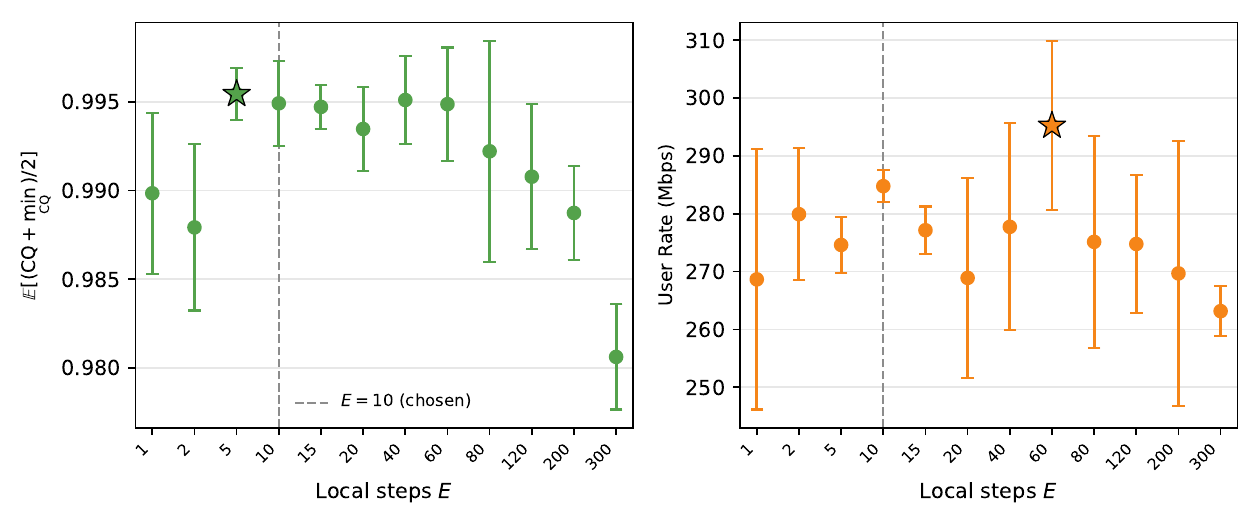}
\caption{Sensitivity of \FedProx{} ($\mu=0.01$) to the number of local steps
$E$ over the dense grid
$\{1,2,5,10,15,20,40,60,80,120,200,300\}$.
Left: composite channel quality
$\mathbb{E}[(\mathrm{CQ}+\min_{\mathrm{CQ}})/2]$. Right: User rate [Mbps]. Star shaped markers highlight the pick value in each plot, while the dashed vertical line indicates the selected operating point ($E=10$).}
\label{fig:e_sweep}
\end{figure}

We next examine the number of local update steps $E$ performed between
consecutive model aggregations. Fig.~\ref{fig:e_sweep} evaluates
$E\in\{1,2,5,10,15,20,40,60,80,120,200,300\}$ for \FedProx{} with
$\mu=0.01$. For each value of $E$, the final global model is evaluated in
terms of composite channel quality and Shannon throughput. The results show
that neither very small nor very large values of $E$ provide the best
performance across the two measures. In particular, $E=10$ provides a
favorable operating point and is therefore used in the remaining experiments.

%------------------------------------------------------------------------------
\subsection{Performance Comparison}
\label{subsec:exp_paper}
%------------------------------------------------------------------------------

We compare the two implementations of the proposed framework, FASCIA-FA and
FASCIA-FP, with three reference methods. The centralized CTDE implementation
of \CARLTON{}~\cite{cohen2024carlton} serves as the learning benchmark, while
Jamming Avoidance Response (JAR) and the Random Agent (RA) provide distributed references. JAR~\cite{cohen2024carlton} switches to an adjacent
channel only when the change improves the channel quality by at least $0.05$,
whereas RA~\cite{cohen2017distributed} selects an initial channel and retains
it throughout the scenario. All methods are evaluated over the same $420$
games, comprising $30$ independent games for each
$N\in\{2,\dots,15\}$.

\begin{figure}[!t]
\centering
\includegraphics[width=\columnwidth]{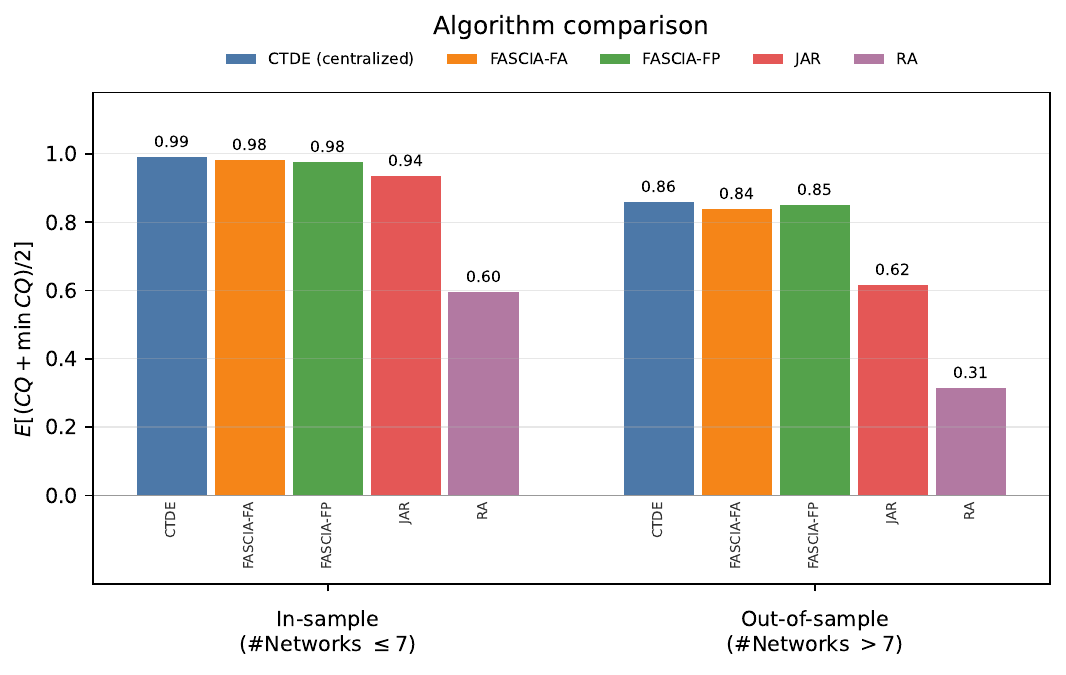}
\caption{In-sample and out-of-sample composite channel quality. The in-sample
results average over network densities represented during training
($N\leq7$), whereas the out-of-sample results average over unseen network
densities ($N>7$).}
\label{fig:fig15_bars}
\end{figure}

\begin{figure}[!t]
\centering
\includegraphics[width=\columnwidth]{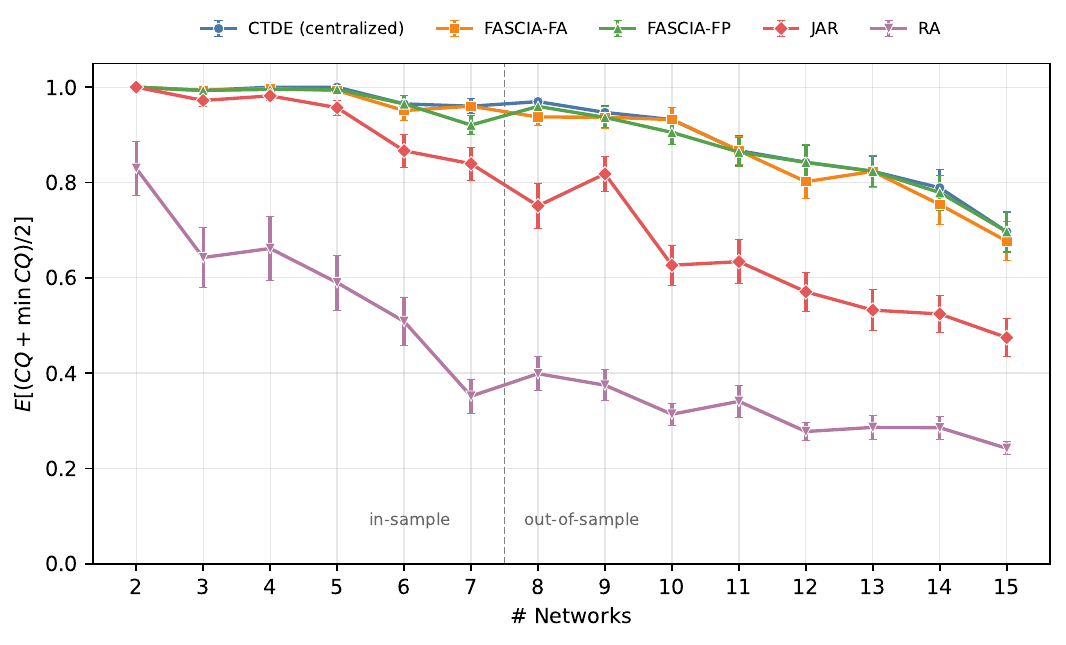}
\caption{Composite channel quality versus the number of networks. The dashed
line separates network densities represented during training ($N\leq7$) from
unseen network densities ($N>7$).}
\label{fig:fig13_curves}
\end{figure}

Fig.~\ref{fig:fig15_bars} first provides an aggregate comparison by separating
the evaluation into two network-density regimes. The in-sample regime contains
$N\leq7$, corresponding to network densities represented during training. In
this regime, CTDE achieves an average composite channel quality of approximately
$0.99$, while FASCIA-FA and FASCIA-FP both achieve approximately $0.98$.
Thus, both federated implementations closely approach the centralized learning
benchmark over the range of network densities represented during training.
JAR achieves approximately $0.94$, whereas RA obtains $0.60$.

The right group of bars in Fig.~\ref{fig:fig15_bars} evaluates generalization
to the unseen network-density regime $N>7$, up to $N=15$. The average
composite channel quality is approximately $0.86$ for CTDE, $0.84$ for
FASCIA-FA, and $0.85$ for FASCIA-FP. Although performance decreases relative
to the in-sample regime as more networks compete for the same $K=10$
channels, both federated implementations remain close to the centralized
benchmark. In particular, FASCIA-FP is within approximately $0.01$ of CTDE
when averaged over the unseen network densities. The corresponding values for
JAR and RA decrease to approximately $0.62$ and $0.31$, respectively,
indicating a substantially larger degradation as the network density
increases.

Fig.~\ref{fig:fig13_curves} complements these aggregate results by reporting
the composite channel quality separately for every network size
$N\in\{2,\dots,15\}$. The vertical dashed line separates the network densities
represented during training ($N\leq7$) from those not encountered during
training ($N>7$). Within the training regime, CTDE, FASCIA-FA, and FASCIA-FP
exhibit very similar performance, with composite channel quality remaining
close to one for the smaller network configurations. As $N$ increases beyond
the training range, the performance of all three learning-based methods
gradually decreases. Nevertheless, both FASCIA implementations continue to
closely track CTDE throughout the out-of-sample range, demonstrating that the
federated model generalizes to network densities not represented during
training.

The per-network-density results also clarify the increasing separation from JAR and RA algorithms. JAR remains below FASCIA over
the evaluated range, with the gap generally increasing for denser configurations, while RA exhibits a substantially larger degradation. At
$N=15$, for example, the composite channel quality remains approximately
$0.70$ for CTDE and the two FASCIA implementations, compared with
approximately $0.47$ for JAR and $0.24$ for RA. Hence, the aggregate
out-of-sample results in Fig.~\ref{fig:fig15_bars} are not driven by only a
small subset of network sizes; Fig.~\ref{fig:fig13_curves} shows that the
federated implementations remain close to the centralized benchmark across
the full range of unseen network densities.

\begin{table}[!t]
\centering
\caption{Average convergence-time score and weighted-score performance}
\label{tab:cts_ws}
\footnotesize
\begin{tabular}{lcc}
\toprule
\textbf{Method} & \textbf{CTS} & \textbf{WS} \\
\midrule
CTDE       & $0.933$ & $0.931$ \\
FASCIA-FA  & $0.933$ & $0.930$ \\
FASCIA-FP  & $0.933$ & $0.930$ \\
JAR        & $0.945$ & $0.910$ \\
RA         & $0.960$ & $0.818$ \\
\bottomrule
\end{tabular}
\end{table}

Table~\ref{tab:cts_ws} complements the channel-quality results by comparing
the convergence-time score (CTS) and weighted score (WS). FASCIA-FA, FASCIA-FP and CTDE achieve a CTS of approximately $0.933$, indicating essentially comparable convergence behavior. JAR and RA attain higher CTS values of $0.945$ and $0.960$, respectively, and therefore stabilize somewhat faster. However, this advantage does not translate into better overall spectrum-coordination
performance. FASCIA-FA and FASCIA-FP both achieve a WS of $0.930$, nearly
identical to the $0.931$ obtained by CTDE and higher than the $0.910$ and
$0.818$ obtained by JAR and RA, respectively.

Taken together, Figs.~\ref{fig:fig15_bars}--\ref{fig:fig13_curves} and
Table~\ref{tab:cts_ws} show that FASCIA preserves performance close to the
centralized CTDE benchmark across both network densities represented during
training and denser configurations not encountered during training. This
behavior is achieved while retaining locally generated interaction experience
at the network managers and coordinating learning through federated model
aggregation rather than centralized experience collection.

%==============================================================================
\section{Conclusion}
\label{sec:conclusion}
%==============================================================================

We presented \OURS{}, a federated agentic AI framework for
interference-aware adaptive spectrum coordination. In \OURS{}, each network
manager operates as an autonomous AI agent that learns from locally generated
interaction experience and makes its spectrum decisions locally, while
coordination across agents is achieved through periodic federated model
aggregation. By replacing centralized experience collection with persistent
local replay and model-level coordination, \OURS{} enables collaborative
learning across interference-coupled networks without pooling their locally
generated interaction experience.

The proposed framework combines Mellowmax-based local value learning with
federated training and proximal regularization to accommodate heterogeneous
and interference-coupled local experience. The resulting FASCIA-FA and
FASCIA-FP implementations achieve spectrum-coordination performance close to
the centralized CTDE benchmark while consistently outperforming existing distributed methods. Importantly, this behavior extends beyond the
network densities represented during training, demonstrating that the learned
federated model can generalize to denser interference environments while
maintaining effective decentralized spectrum decisions.

Overall, the results demonstrate that federated model-level coordination can
provide an effective alternative to centralized training for adaptive spectrum
coordination. \OURS{} retains the benefits of collaborative learning while
allowing interaction experience to remain at the individual network managers
and preserving decentralized execution. These findings support federated
agentic AI as a promising approach for scalable and adaptive spectrum
coordination in distributed, interference-coupled wireless networks.

%==============================================================================

\bibliographystyle{IEEEtran}
\bibliography{refs}

\end{document}